\documentclass[useAMS,usenatbib]{mn2e}
\usepackage{psfig}
\usepackage{amssymb}

\title[The SDSS DR6 Luminosity Functions of Galaxies]
{The SDSS DR6 Luminosity Functions of Galaxies} 

\author[Antonio D. Montero-Dorta \&
Francisco Prada
 ]{
\parbox[t]{\textwidth}{
Antonio D. Montero-Dorta$^1$ \&
Francisco Prada$^1$}
\vspace*{6pt} \\ 
$^1$Instituto de Astrof\'isica de Andaluc\'ia (CSIC), Granada, E-18008, Spain
\vspace{-0.5cm} 
}

\date{Accepted ---. Received ---;in original form ---}

\pubyear{2005}

\newcommand{\plotone}[1]
           {\centering \leavevmode \psfig{file=#1,width=\columnwidth,clip=}}

\def\simlt{\lower.5ex\hbox{$\; \buildrel < \over \sim \;$}}
\def\simgt{\lower.5ex\hbox{$\; \buildrel > \over \sim \;$}}
\usepackage{graphicx}
\usepackage{rotating}

\begin{document}

\bibliographystyle{mnras}

\maketitle


\begin{abstract}

We present number counts, luminosity functions (LFs) and luminosity densities
of galaxies obtained using the Sloan Digital Sky Survey Sixth Data Release in all
SDSS photometric bands. Thanks to the SDSS DR6, galaxy statistics have increased by a factor of $\sim9$
in the $u$-band and by a factor of $\sim4-5$ in the rest of the SDSS bands
with respect to the previous work of \cite{Blanton2003a}. In addition,
we have achieved a high redshift completeness in our galaxy samples. Firstly, by 
making use of the survey masks, provided by the NYU-VAGC DR6, we have been able to define 
an area on the sky of high angular redshift completeness. Secondly,
we guarantee that brightness-dependent redshift incompleteness
is small within the magnitude ranges that define our galaxy samples. With these advances, we have estimated very accurate SDSS DR6 LFs
in both the bright and the faint end. In the $^{0.1}r$-band, our SDSS DR6 luminosity function
is well fitted by a Schechter LF with parameters  
$\Phi_{*}=0.90 \pm 0.07$, $M_{*}-5log_{10}h=-20.73 \pm 0.04$ and $\alpha=-1.23 \pm 0.02$. 
As compared with previous results, we find some notable differences.
In the bright end of the $^{0.1}u$-band luminosity function we
find a remarkable excess, of $\sim1.7$ dex at $M_{^{0.1}u}\simeq-20.5$,  with respect to the best-fit Schechter LF. This excess weakens in the $^{0.1}g$-band, fading away towards
the very red $^{0.1}z$-band. A
preliminary analysis on the nature of this bright-end bump reveals that it
is mostly comprised of active galaxies and QSOs. It seems, therefore, that an important fraction
of this exceeding luminosity may come from nuclear activity. In
the faint end of the SDSS DR6 luminosity functions,
where we can reach $1-1.5$ magnitudes deeper than
the previous SDSS LF estimation, we obtain a steeper slope, that
increases from the $^{0.1}u$-band, with $\alpha=-1.01 \pm 0.03$, to the very red $^{0.1}z$-band, with
$\alpha= -1.26 \pm 0.03$. 
These state-of-the-art results may
be used to constrain a variety of aspects of star formation histories and/or
feed-back processes in galaxy formation models.
   
\end{abstract}


\section{Introduction}
\label{sec:intro}

From the pioneering work of \cite{Humason1956} and \cite{Sandage1978},
who measured redshifts of bright galaxies from the Shapley-Ames 
photometric catalog \cite{Shapley1932}, much effort 
has been invested in mapping the luminous and matter contents of the Universe. The 
Center for Astrophysics survey (CfA, \citealt{Huchra1983}) is
considered as the first proper redshift survey,  
specifically designed for cosmology studies.
More important, from the first slice of about $1,000$ 
galaxies the CfA Redshift Survey provided the community with observational
evidence of an old theoretical, and at times controversial idea:
the existence of a large scale structure of galaxies  in the Universe \citep{Davis1985}. This first vision
of cosmic complexity encouraged the development of new imaging 
and spectrometric technology and, consequently,
gave rise to a number of other redshift surveys that followed
different approaches and strategies. To name but a few,
the Southern Sky Redshift Survey (SSRS, \citealt{daCosta1988}); 
the Perseus-Pisces catalog \citep{Giovanelli1991} or the
catalogs based on data from the Infrared Astronomical Satellite
(IRAS). In the last decade, the emergence of multi-fiber
spectrographs set the scene for larger and deeper redshift surveys. Examples of these are 
the Las Campanas Redshift Survey (LCRS, \citealt{Shectman1996}), 
consisting of 26,418 galaxies with an average redshift of $z\sim0.1$; or
the 2 degree Field Redshift Survey (2dF RS, \citealt{Colless2001}), 
with about $222,000$ galaxies and  covering a sky area of $1500 ~deg^{2}$.
Finally, the Sloan Digital Sky Survey (SDSS, \citealt{York2000}) is
the largest photometric and spectroscopic survey ever compiled, and represents
the most accurate map of the nearby universe at $z\lesssim0.3$.
The SDSS Sixth Data Release \citep{Adelman2008}, 
that we use in this paper, contains spectroscopic information 
for more than 1,000,000 galaxies and quasars which spread over 
$7425 ~deg^{2}$ on the sky. Only in recent years, with surveys like 
the DEEP2 Galaxy Redshift Survey \citep{Davis2003} or the
VIMOS-VLT Deep Survey (VVDS, \citealt{LeFevre2003}), have we
reached the stage where we can study the 
galaxy population in the distant ($z\sim1$) Universe. Other high-z
surveys are currently being completed.  

The advances in the survey field also made it necessary to develop
data reduction pipelines and 
analysis tools to process and understand increasingly larger
data sets. These days, cosmologists use a number of 
statistics to characterize, for a particular survey,
the distribution of galaxies in three-dimensional space. Number 
counts, selection functions, luminosity functions or correlation
functions are just a few examples. In this work we focus on the number
counts and the luminosity functions of galaxies, that
we draw from the SDSS DR6. Number counts,
which describe the distribution of fluxes of galaxies, have been
calculated for a number of surveys. The general consensus is that, 
in the close-by universe, galaxy number counts look like what we expect from an
euclidean, not-evolved Universe. \cite{Yasuda2001} obtained number counts for the SDSS 
Commissioning Data in all ugriz bands. \cite{Norberg2002} provided number counts for the
2dF survey in the $b_{j}$ band. \cite{Feulner2007} not only 
estimated galaxy number counts for a set of catalogs based 
on the Munich Near-Infrared Cluster Survey (MUNICS, \citealt{Drory2001})
but also presented a complete revision of this subject in the literature
(see their Figure 8). In contrast to galaxy number 
counts, the luminosity function (LF), which is the number
density of galaxies per unit absolute magnitude, 
has been historically a rather controversial issue. For example, \cite{Marzke1994}
- using the CfA RS -,  \cite{Norberg2002} 
- 2dF RS - and 
\cite{Blanton2003a} - SDSS DR2-, all obtained very different results.

Both the luminosity function and the luminosity density of galaxies
are observational signs of the process of galaxy formation and evolution. 
A precise determination of these statistics is needed to 
constrain current theories. Consequently, new discoveries
in observational cosmology could make a strong 
impact in our understanding of the physical processes that drive the 
birth and life of galaxies in the Universe. Nowadays, state-of-the-art models
of galaxy formation invoke a number of galactic "mechanisms", which are connected
through the so-call feed-back processes. Disentangling
these relations is an ambitious but crucial task in modern Cosmology. 
In this sense, the semi-analytic models of galaxy formation (SAMs, e.g. \citealt{Croton2006}),
which are embedded in N-body simulations like the Millennium Run
(see \citealt{Springel2005}), are a very useful tool for cosmologists. These SAMs 
are a good ground for testing new theoretical ideas and understanding their 
observational implications.

The main purpose of this work is to take advantage of the large
increase in the galaxy statistics thanks to the SDSS DR6 to obtain the
number counts, LFs and luminosity densities
of galaxies in the nearby universe. We intend 
to shed light both in the faint end of the LF, where most discrepancies
come from, and in the bright-end, where statistics have always been poor 
and errors, consequently large. In section~\ref{sec:dr6} we 
briefly describe the SDSS DR6, discuss our sample selection
and comment on redshift completeness. In section~\ref{sec:results} we present
our results on the number counts,
the LFs and the luminosity densities of galaxies
in each one of the SDSS photometric bands. Finally, in Section~\ref{sec:discussion} 
we discuss our results and in Section~\ref{sec:sum} we present a summary of our work.
Throughout this paper, unless otherwise stated, we assume a standard $\Lambda$CDM
concordance cosmology, with $\Omega_m=0.3$, $\Omega_\Lambda=0.7$,
$w=-1$, and $h=1$. In addition, we use AB magnitudes.

\section{The Sloan Digital Sky Survey Data Release 6. Data samples selection and redshift completeness}
\label{sec:dr6}

In this work, we use the SDSS Sixth Data Release 
\citep{Adelman2008}. This data set completes the North
Galactic Cap, containing photometric information of $\sim290$ million
objects over 9583 $deg^{2}$. Around $1.27$ million objects were selected
for spectroscopy, covering an area of $7425 ~deg^{2}$ on the sky. Important 
for this work, spectroscopy is available for $\sim1,000,000$ galaxies down to 
magnitude $r\sim17.77$ \citep{York2000, Stoughton2002}. Detailed
information about the SDSS DR6 can be found in 
\cite{Adelman2008}.

The SDSS DR6 is the largest survey of the nearby universe
publicly available. The SDSS collaboration have successively extended
their catalogs since the times of the SDSS Early Data Release and, 
consequently, improved enormously our capability of mapping the universe up to
redshift $z\sim0.25$. A few years ago, \cite{Blanton2003a} used the SDSS DR2
to estimate the luminosity function of galaxies. In this work,
thanks to the SDSS DR6, the size of our samples has
risen by a factor $\sim9$ in the very blue u-band
and by a factor $\sim4-5$ in the other bands ( g, r, i and z).
This huge enhancement in the statistics will be especially 
critical in the bright end of the LF, 
where statistics have always been poor in the past. In addition,
we also expect to reach deeper magnitudes in the
faint end with respect to previous works. Galaxy number counts can also
be estimated with significantly more accuracy. 
It is therefore well justified to update 
the current knowledge on the number counts, luminosity functions and luminosity densities of 
galaxies in the close-by Universe.

\subsection{Data samples selection}
\label{sec:selection}

In this section, we describe the selection of our
galaxy samples. We have drawn our samples from the NYU Value Added
Galaxy Catalog DR6  Large Scale Structure sample \citep{Blanton2005b}. The NYU-VAGC
is a compilation of galaxy catalogs cross-matched to the SDSS that 
includes a number of useful quantities derived from the photometric
and the spectroscopic catalogs (such as K-corrections or absolute
magnitudes). It also incorporates a precise and user-friendly description of the geometry 
of the survey, which has never been used in previous SDSS LF works.

\begin{figure*}
\begin{center}
\begin{tabular}{c}
\includegraphics[scale=0.7]{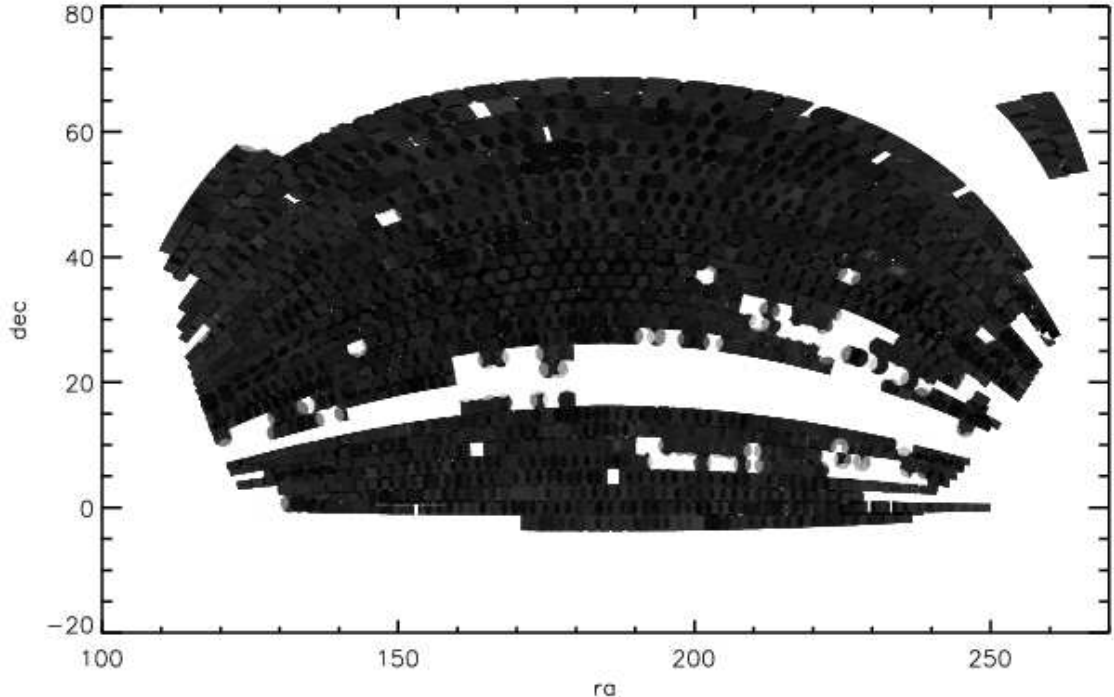}
\end{tabular}
\end{center}
\caption{Angular redshift completeness of the SDSS DR6 spectroscopic catalog in the right ascension range 
$110^{\circ}< RA < 270^{\circ}$, that encompasses $\sim99\%$ of the survey. Each
polygon in the plot is an area of constant completeness. Polygons are 
also color-coded in white to black tones, meaning from $0\%$ to
$100\%$ completeness.}
\label{fig:masks}
\end{figure*}

Before we proceed with the sample selection, we will describe how we built our
Parent Sample from the NYU-VAGC LSS sample. The Parent Sample is the galaxy catalog 
from which we extract all samples that we use in this work. Firstly,
we take all objects that satisfy the criteria
of the SDSS Main Galaxy Sample \citep{Strauss2002}.
 Secondly, we make use of the survey
masks, provided by the NYU-VAGC, to define our {\sl{high-completeness
survey window}} (hereafter, HCSW). In this paper we deal 
with number counts and luminosity functions in the entire 
SDSS DR6 spectroscopic catalog. These tasks require knowing the 
effective area covered on the sky by the sample . In practice, 
redshift completeness, which is the fraction of galaxies with a successful redshift measurement, 
is far from being uniform across the sky. In 
Figure~\ref{fig:masks}, we show the angular sky redshift completeness of the SDSS DR6 spectroscopic catalog for a major
part of the survey ($\sim99\%$), just excluding objects outside the 
range $110^{\circ} < RA < 270^{\circ}$, for the sake of clarity.
The plot has been pixelized so that each pixel - or polygon - is an 
area of constant completeness. Polygons are also color-coded 
in white to black tones, the latter meaning $100\%$ completeness.
In order to avoid overestimating the area covered on the sky 
by our samples, the HCSW is defined so that 
redshift completeness is at least $80\%$ in every polygon. Only galaxies
lying in these polygons are included in our Parent Sample.

With these first restrictions our Parent Sample is comprised of $947,053$ galaxies
that spread over $6428.32 ~deg^{2}$ on the sky. From this catalog, 
each sample is drawn by applying the following cuts to the redshift and apparent magnitude:\\

\begin{itemize}
  \item $m_{min}$(j) $<$ m(j) $<$ $m_{max}$(j)
  \item $z_{min}$(j) $<$ z(j) $<$ $z_{max}$(j)
  \end{itemize}

where j=u,  g,  r,  i,  z. In Table~\ref{table:samples} we show lower and upper limits of these quantities
along with the number of galaxies for each sample.

\begin{table}
\begin{center}
\begin{tabular}{cccccc}
\hline
\hline
      Band &      Number&        $m_{min}$ &       $m_{max}$ &  $z_{min}$ &    $z_{max}$ \\
\hline
         u &      192,068 &        16.45 &       19.00 &   0.02&    0.19\\

         g &      241,719&        14.55 &        17.91 &    0.02&    0.16\\

         r &      516,891&        13.93 &        17.77 &    0.02&    0.22 \\

         i &      429,173&        13.55&         17.24&    0.02&    0.22\\

         z &      414,828&        13.40&         16.97&    0.02&    0.23 \\

\hline
\hline
\end{tabular}
\end{center}
\caption{Number of galaxies and limits in apparent magnitude and redshift of each SDSS photometric band sample. 
Motivation for each cut is discussed in section~\ref{sec:selection}. }
\label{table:samples}
\end{table}

At this point, it is convenient to clarify the 
motivation for each cut. The apparent magnitude limits of Table~\ref{table:samples}
are set to ensure that the effect of redshift incompleteness is
small in our SDSS galaxy samples. In the bright 
end, brightness-dependent redshift incompleteness starts to
be important at $r\lesssim15$. In the faint end, redshift incompleteness
in all SDSS bands is dominated by the intrinsic faint limit of the 
Main Galaxy Sample \citep{Strauss2002}, i.e. $r=17.77$. In order 
to choose these apparent magnitude limits, 
we have made use of the SDSS galaxy number counts, that 
will be properly discussed in Section~\ref{sec:counts}. For 
each galaxy sample, we have taken the magnitude range 
where the number of galaxies rises at constant rate in each SDSS band, with a $0.1$-dex 
deviation allowance (see Figure 4, where galaxy number counts 
have been scaled by an euclidean, not-evolved model). Within these limits, 
we have estimated that redshift completeness is $\sim85\%$ in all
SDSS bands. In Section~\ref{sec:completeness}, we will discuss on 
redshift incompleteness issues in the SDSS in more detail. 
 
\begin{figure}
\plotone{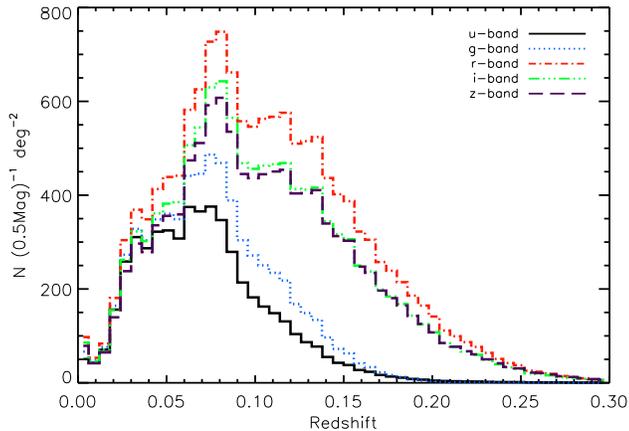}
\caption{Redshift distributions in all SDSS photometric bands. }
\label{fig:z_dist}
\end{figure}

In Figure~\ref{fig:z_dist}, we show the redshift distribution for the Parent Sample in 
each SDSS photometric band.
This figure illustrates the motivation for the
redshift limits given in Table~\ref{table:samples}. The lower redshift limit 
is set to $z=0.02$ to avoid the redshift
incompleteness that affects the very bright and nearby galaxies. The upper redshift 
limit corresponds to the redshift at which $95\%$ of objects are selected 
in each sample and is set for consistency.

\begin{figure}
\plotone{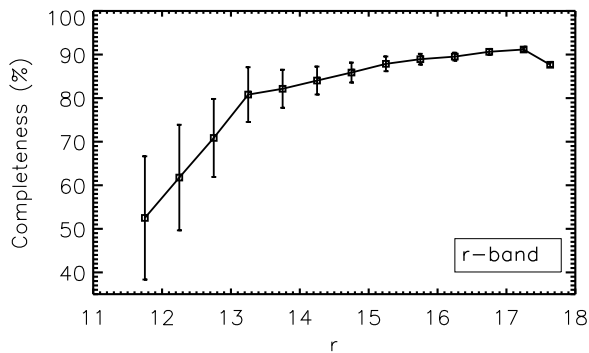}
\caption{Redshift completeness vs. apparent magnitude in the r-band. At $r\lesssim14$
completeness decreases sharply. Errors have been estimated by 
propagating the poissonian uncertainties to the redshift completeness.}
\label{fig:completeness}
\end{figure} 

\subsection{Redshift Completeness}
\label{sec:completeness}

It is well known that the brightest galaxies in the SDSS catalogs
are affected by a severe redshift incompleteness. This effect
is especially important for nearby galaxies. Apparently large
and complex objects represent a major task for the photometric
SDSS pipelines \citep{Strauss2002}. In addition,
and affecting the entire magnitude range, is the so-called {\sl{fiber collisions problem}}. 
This source of incompleteness is due to the
fact that fibers cannot be placed closer than $55''$.
The consequence of this limitation in the SDSS data
 is estimated in a $6\%$ incompleteness by \cite{Strauss2002}.

In Figure~\ref{fig:completeness}, we show 
how redshift completeness varies with apparent magnitude in the r-band.
At $r\simeq14$, redshift completeness falls from $\sim85\%$ to $\sim50\%$ at $r\simeq12$.  
In the faint end, however, it reaches a plateau at $~90\%$ - down to $r\simeq17.8$ -, which is in agreement with \cite{Strauss2002}. In the rest of the SDSS bands, redshift completeness
not only decreases in the bright end, but also in the faint end. This is
due to the intrinsic r-band faint limit of the Main Galaxy 
Sample at $r=17.77$.  Interestingly, the apparent magnitude, $m_{min}(j)$, at 
which this decrease occurs varies between bands as a result
of the dispersion in the colors of galaxies. We have checked that, by imposing 
$m_{max} (j)> m(j) > m_{min}(j)$, we ensure that each 
galaxy sample is approximately $85\%$ complete
in any magnitude bin. 
 
Another aspect of redshift completeness that should be carefully 
taken into account is its angular variation. As we discussed above,
a precise estimation of the area covered in the sky by our samples
is needed in order to calculate number counts and luminosity 
functions of galaxies. We ensure here that we do not overestimate 
this area by defining a region, the HCSW, in which redshift completeness
is at least $80\%$. The HCSW covers $6428.32 ~deg^{2}$, about $90\%$ of the
entire sky coverage of the SDSS DR6 spectroscopic catalog. As we
can see in Figure~\ref{fig:masks}, the SDSS DR6 sky coverage is patched with areas where no redshift
was measured at all. These regions cover a total of $623.5 ~deg^{2}$ on the sky.

Finally, we cannot discard the possibility that the SDSS spectroscopic
catalog is incomplete for very low surface brightness objects, i.e. $\mu_{r,50}\gtrsim24$.
(see \citealt{Strauss2002})
The presence of  this selection effect in the data could in principle affect our 
results. However, we have evidence that the surface
brightness of most galaxies in the SDSS, and consequently in our galaxy samples,
is far greater than $\mu_{r,50}\simeq24$ \citep{Blanton2003a}.

\section{Results}
\label{sec:results}

\subsection{Number counts}
\label{sec:counts}

\begin{figure}
\plotone{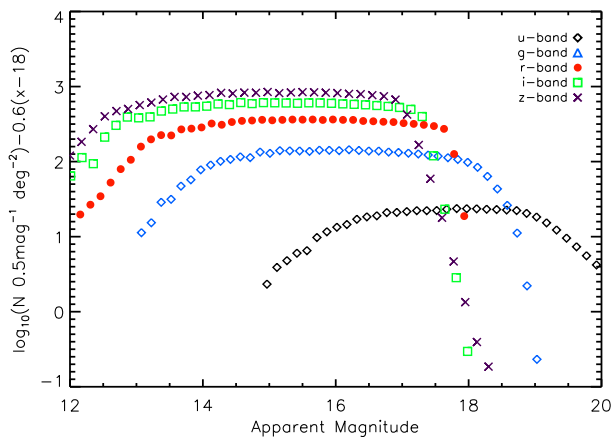}
\caption{Galaxy number counts in all SDSS bands scaled by an euclidean model 
in bins of half a magnitude. 
Poissonian errors are of similar size as symbols, so they
are not shown.}
\label{fig:counts}
\end{figure}

In Figure~\ref{fig:counts} we plot with different symbols 
the logarithm of the number of galaxies per unit 
area and apparent magnitude (actually, half magnitude), 
scaled by an euclidean model for all SDSS bands. We choose,
arbitrarily, the following model for euclidean counts:

\begin{equation}
N_{euclidean}=10^{0.6(x-18)}
\label{eqn:euclidean}
\end{equation}  
where $x$ is replaced by the apparent magnitude in each SDSS band.    

In the r-band, galaxy number counts increase by a factor of
about $10$ from $r\sim12$ to $r\sim13.5$. This
magnitude range is strongly affected by redshift incompleteness, as discussed above
(see also Figure~\ref{fig:completeness}). From $r\sim14$
to $r\sim18$, counts rise at 
approximately the same rate as that of the model 
(Equation~\ref{eqn:euclidean}). At $r\sim18$, where the SDSS spectroscopic faint-end limit is set
($r=17.77$), galaxy number counts fall sharply.

In the rest of the bands, the behavior 
is very similar in the bright end and it is also due to redshift incompleteness.
In the z-band, counts start to follow the euclidean model at $z\sim13.5$.
In the i-band, this happens at $i\sim13.5$; in the
g-band, at $g\sim14.5$ and, in the u-band, at $u\sim16.5$.
In the faint end, galaxy number counts fall at $u\sim19$,
$g\sim18$, $i\sim17$ and $z\sim17$. This 
decrease is obviously less pronounced than in the r-band.
This reflects the fact that the color-magnitude relation
is not univoque. In addition, the dispersion in this relation
is considerably larger for $u-r$ colors
than it is for $r-z$ colors 
\citep{Blanton2003b}. As seen 
in Figure~\ref{fig:counts}, the slope of the
faint-end decrease in the galaxy number counts is
considerably steeper in the u and g-bands than it
is in the redder bands.   

Number counts are consistent with an Euclidean, not-evolved Universe in 
all SDSS bands within the magnitude and redshift ranges given in Table~\ref{table:samples}.
Within these ranges, the slopes in units of $mag^{-1}$ of the SDSS DR6 galaxy number counts are
listed in Table~\ref{table:counts}. Although discrepancies exist with respect to the euclidean model
, these are probably smaller than the expected uncertainty in the determination 
of the apparent magnitude and redshift of each galaxy in the SDSS.

\begin{table}
\begin{center}
\begin{tabular}{cccc}
\hline
\hline
      Band &      Apparent magnitude limits&    Redshift limits&    Slope \\
\hline
         u &      $16.45 < u < 19.00$ &   $0.02<z<0.19$&   0.627\\

         g &      $14.55 < g < 17.91$ &    $0.02<z<0.16$&         0.608\\

         r &      $13.93 < r < 17.77$ &   $0.02<z<0.22$&    0.610\\

         i &      $13.55 < i < 17.24$&      $0.02<z<0.22$&     0.611\\

         z &      $13.40  < z < 16.97$&          $0.02<z<0.23$&      0.626\\

\hline
\hline
\end{tabular}
\end{center}
\caption{Slope in units of $mag^{-1}$ of galaxy number counts
within the apparent magnitude and redshift limits of each sample. Within these ranges, 
number counts are consistent with an Euclidean, not-evolved Universe.} 
\label{table:counts}
\end{table}

Our results for the nearby universe are in agreement with a number of
previous works (see \citealt{Feulner2007} for a review). However, only \cite{Yasuda2001}
obtained galaxy number counts for SDSS data. The authors used 
imaging data taken only during the commissioning phase of the SDSS. They found 
an euclidean-like behavior up to magnitude $m_{min}\sim12$ - except
for the u-band, where $m_{min}\sim14$ - (see their Figure 8). The
lack of galaxies in the very bright end of Figure~\ref{fig:counts} with respect to photometric
number counts from \cite{Yasuda2001} is partially due to the 
strong redshift incompleteness that affects the SDSS spectroscopic catalog
in these magnitude ranges (see Section~\ref{sec:completeness}). However, our results
and those from \cite{Yasuda2001} are not directly comparable.
Firstly, they used a very limited sample, in terms of sky coverage ($\sim230~deg^{2}$).
Secondly, comparing spectroscopic and photometric results
is always tricky.

\subsection{Luminosity Functions}
\label{sec:LF}

\begin{figure}
\plotone{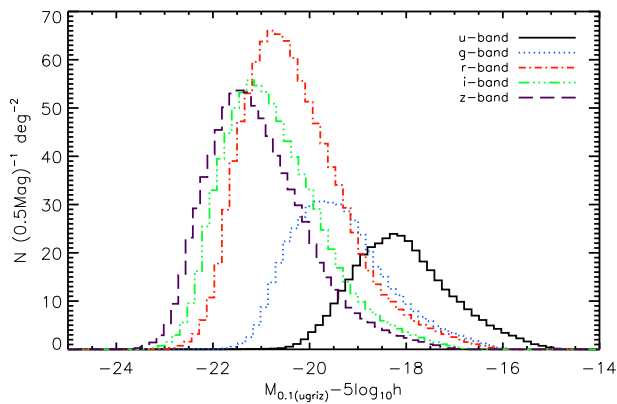}
\caption{Absolute magnitude distribution in each sample. Although the shape 
of the distributions is very similar, mean values 
move towards bright-end magnitude bins from $^{0.1}u$-band to $^{0.1}z$-band, as red objects are, 
on average, brighter than blue objects.}
\label{fig:absmag}
\end{figure}

\begin{figure}
\plotone{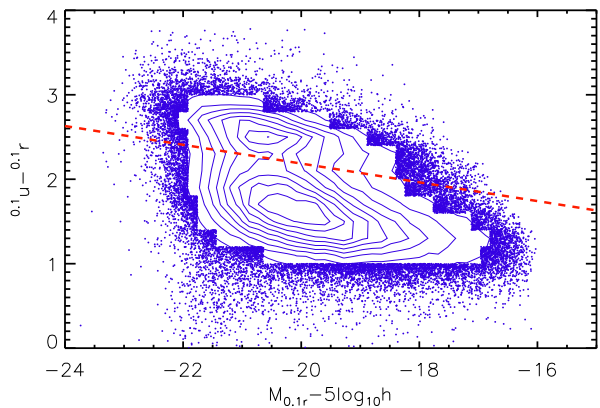}
\caption{The $(^{0.1}u-^{0.1}r)$ color-magnitude diagram in the $^{0.1}r$-band.
 The dashed line represents the 
demarcation commonly used to separate red and blue galaxies.}
\label{fig:cmd}
\end{figure}

In order to estimate the luminosity function of galaxies in each SDSS photometric band, we take
absolute magnitudes and K-corrections from 
the NYU VAGC DR6 LSS catalog. Following \cite{Blanton2003a}, absolute
magnitudes are calculated with the SDSS photometric bands 
shifted to $z=0.1$. With this convention, the absolute magnitude of 
a galaxy in a given band shifted to $z=0.1$, $M_{^{0,1}j}$, would be constructed from 
its apparent magnitude at $z=0$, $m_{j}$, and its redshift z as follows:

\begin{equation}
M_{^{0,1}j}=m_{j}-5log_{10}h-DM(z)-K_{^{0,1}j}(z)
\label{eqn:mag}
\end{equation} 

where $DM(z)$ is the distance modulus (which depends also on the cosmological
parameters) and $K_{^{0,1}j}(z)$, the K-correction for the galaxy in the shifted band $^{0.1}j$. 
 \cite{Blanton2003a} included another correction in Expression~\ref{eqn:mag} to 
 account for the evolution of the luminosity of a galaxy from redshift z to $z=0$, the
 so-called evolution correction. However, we expect this correction to be very small
 within the redshift ranges that we have considered here (see
 also Section~\ref{sec:discussion}). In Figure~\ref{fig:absmag}, 
 we show the distribution of 
K-corrected absolute magnitudes in each galaxy sample.  The shape of the 
absolute magnitude distribution is very similar in all SDSS bands: a gaussian-like 
distribution slightly skewed to fainter magnitudes. However, mean values move towards 
bright-end bins from $^{0.1}u$-band - $M_{^{0.1}u}\sim-18$ -  to $^{0.1}z$-band - 
$M_{^{0.1}z}\sim-21.5$, 
which is consistent with the fact that red objects are, on average, 
brighter than blue objects. In Figure~\ref{fig:cmd}, we also show the bimodal 
$^{0.1}u-^{0.1}r$ color distribution of galaxies in our $^{0.1}r$-band sample. With
a dashed line we represent the demarcation commonly used 
to separate red and blue objects (see \citealt{Strateva2001}).

      We use the Stepwise 
Maximum Likelihood Method (SWML; \citealt{Efstathiou1988})
to estimate the LF of galaxies, that is commonly expressed as $\Phi(L)$. This technique
does not rely on any assumption about the shape of $\Phi(L)$ and is 
considered as the most reliable estimator of the LF. For 
more details about this method see \cite{Efstathiou1988} and \cite{Norberg2002}. The 
SWML requires an independent estimation of the number density of 
galaxies, i.e. the normalization constant, $n$. We use the following 
prescription proposed by \cite{Davis1982} based on the selection
function of each galaxy, $\phi(z_{i})$, and the maximum volume encompassed
by the sample, $V_{max}$:

\begin{equation}
n=\frac{1}{V_{max}}\sum_{i}\frac{1}{\phi(z_{i})}
\label{eqn:norm}
\end{equation}

\begin{figure}
\plotone{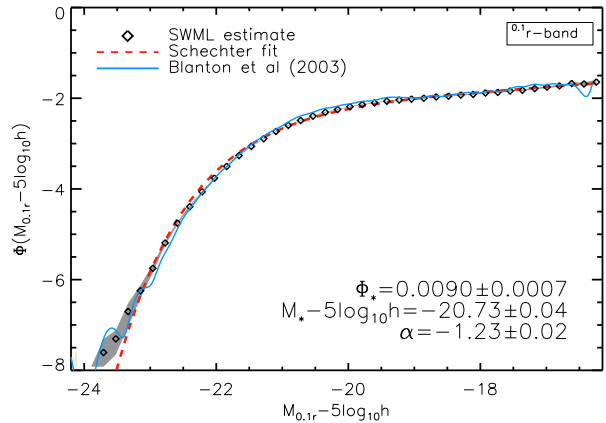}
\caption{The $^{0.1}r$-band SDSS DR6 Luminosity Function. The SWML LF estimate is shown
in diamonds. The dashed line represents the best-fit Schechter
function and the solid line, the $^{0.1}r$-band LF from Blanton et al. (2003b). 
Best-fit values of Schechter parameters $\alpha$, $M_{*}$ and $\Phi_{*}$ are
also shown in the figure. Shaded regions represent the 1$\sigma$ 
uncertainty calculated using a bootstrapping technique.}
\label{fig:lf_r}
\end{figure}

In Figure~\ref{fig:lf_r} we show the SWML estimate of the
SDSS DR6 LF in the $^{0.1}r$-band. In addition, we
over-plot with a dashed line the best-fit Schechter function
\citep{Schechter1976}, which has the following shape:

\begin{eqnarray*}
\Phi(M)=0.4\log(10)\Phi_{*}10^{-0.4(M-M_{*})(\alpha+1)}\\
\exp\big[-10^{-0.4(M-M_{*})}\big]~~~~~~~~~~~~(4)
\label{eqn:sch}
\end{eqnarray*}

where $\alpha$, $M_{*}$ and $\Phi_{*}$ are the three parameters to fit.
Values of these parameters for the best-fit Schechter function are 
given in Table~\ref{table:LF}. For comparison, we also show in Figure~\ref{fig:lf_r}  the LF of  \cite{Blanton2003a} with a solid line. This comparison will
be addressed in the Discussion section. 
To calculate errors in the SWML
estimates of the LF we perform a bootstrapping
analysis using $1,000$ random sub-samples of $1/3$ of the 
number of objects in each sample. In Figure~\ref{fig:lf_r}, shaded 
regions represent the $1\sigma$ uncertainty obtained from this method. 

Due to the big number statistics that we have, with about
$500,000$ galaxies in the $^{0.1}r$-band, errors are only significant 
in the very bright end of the LF. In the faint end, we can go down to 
$M_{^{0.1}r}\sim-16$, which means that we can build the LF 
with unprecedented precision within a very large range of magnitudes.
As we will see below, the above statements hold for all SDSS bands.
Our $^{0. 1}r$-band LF is reasonably well fitted by a Schechter LF
with a faint-end slope $\alpha=-1.23$. It is only in the 
very bright end where this 
best-fit Schechter LF starts to underestimate our LF. At
$M_{^{0.1}r}\lesssim-23.5$, statistics are  
poor and errors become increasingly large.

\begin{figure*}
\begin{center}
\begin{tabular}{c}
\includegraphics[scale=0.7]{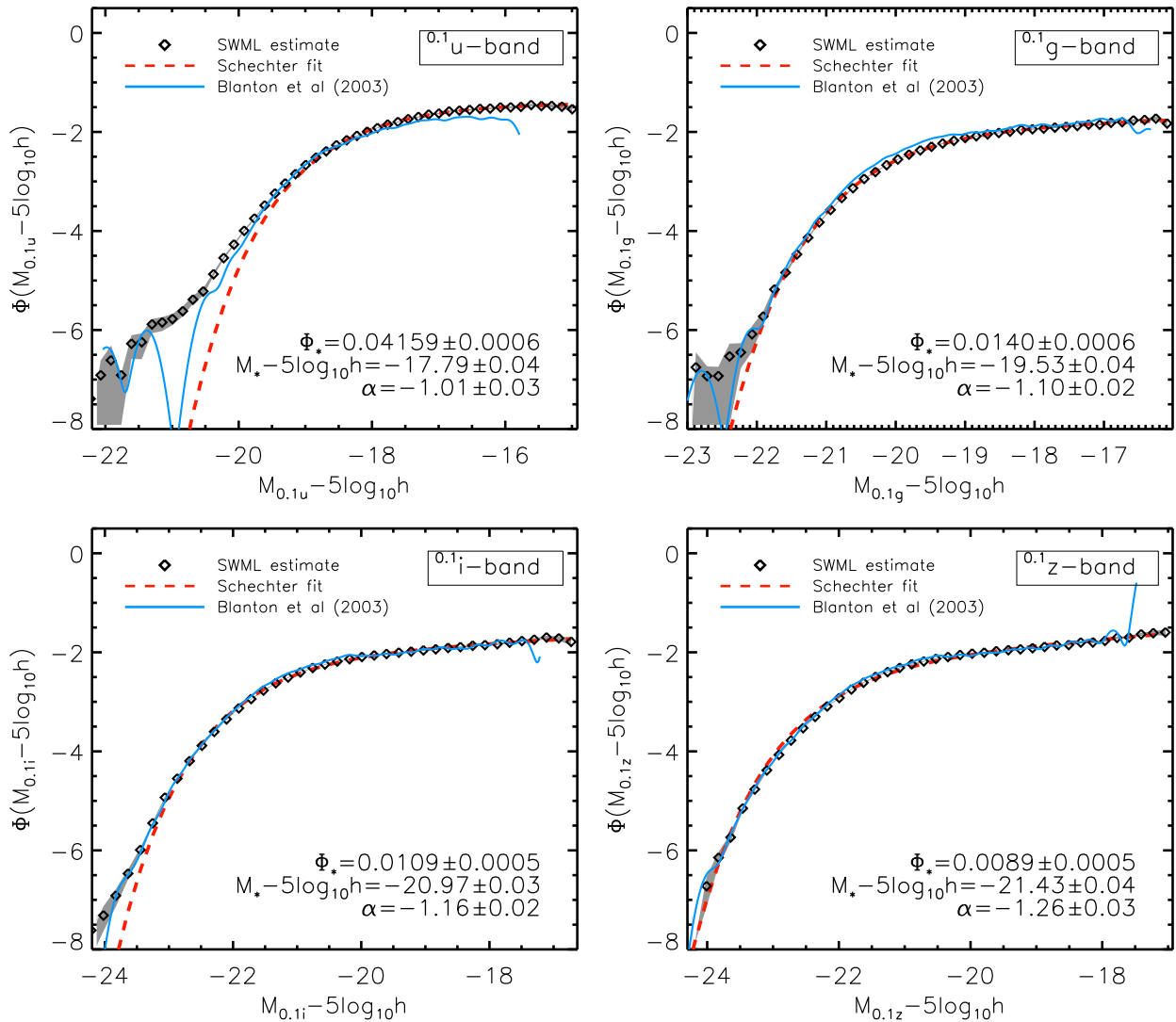}
\end{tabular}
\end{center}
\caption{The SDSS DR6 Luminosity Functions in bands $^{0.1}u$, $^{0.1}g$, $^{0.1}i$ and $^{0.1}z$. 
SWML LF estimates 
are shown in diamonds and best-fit Schechter
functions are represented by dashed lines. 
In addition, we over-plot in each panel the LF from Blanton et al. (2003b).
Best-fit values of 
Schechter parameters $\alpha$, $M_{*}$ and $\Phi_{*}$ are
also shown. Shaded regions represent the 1$\sigma$ 
uncertainty calculated using a bootstrapping technique.}
\label{fig:ugiz_lf}
\end{figure*}

In Figure~\ref{fig:ugiz_lf} we present, in the same way as in Figure~\ref{fig:lf_r},
SWML estimates of the LF in bands $^{0.1}u$, $^{0.1}g$, $^{0.1}i$ and $^{0.1}z$, as well as their corresponding best-fit Schechter LF. Values
of best-fit Schechter parameters are also given in Table~\ref{table:LF}.
As in the $^{0.1}r$-band, errors are only significant in the
very bright end of $^{0.1}u$, $^{0.1}g$, $^{0.1}i$ and $^{0.1}z$ - band luminosity functions.
In addition, we can go down to very faint magnitudes
without losing precision.

In the very blue $^{0.1}u$-band, the shape of our SDSS DR6 LF is
consistent with a Schechter LF with an almost flat 
faint-end slope (corresponding to $\alpha=-1.01$). However, in the
bright end, we find a remarkable excess with respect to the 
best-fit Schechter LF. This excess, of $\sim1.7$ dex at $M_{{0.1}^u}\eqsim-20.5$, is very significant 
within the magnitude range $-20.5<M_{^{0.1}u}\lesssim-22$. In
the $^{0.1}g$-band, the bright-end bump weakens considerably, 
but it is probably still significant, even though errors are large 
according to our bootstrapping analysis. In this band, our SDSS LF is 
very well fitted by a Schechter LF with a positive faint-end slope, corresponding 
to $\alpha=-1.10$. Only in the very bright-end, where the excess is
still noticeable, do we find some discrepancy.   
Below, we provide a preliminary analysis 
and discussion on the nature of the $^{0.1}u$-band LF bright-end bump, that may
have important implications in terms of galaxy formation and evolution.

In the redder bands we find a positive
faint-end slope, corresponding to $\alpha=-1.16$ in the $^{0.1}i$-band and $\alpha=-1.26$
in the $^{0.1}z$-band. The bright-end bump has diminished substantially 
in the $^{0.1}i$-band and has disappeared in the very red $^{0.1}z$-band.
It is interesting to note that, from the $^{0.1}u$-band to the $^{0.1}z$-band, the shape of the SWML estimate of
the SDSS LF changes following a clear pattern. The faint-end slope
increases towards the redder bands (see Table~\ref{table:LF}), being almost
flat in the $^{0.1}u$-band and remarkably steep in the $^{0.1}z$-band. Conversely,
the excess of bright galaxies with respect to the best-fit Schechter function
diminishes towards redder bands. In principle, one might think that the strong
bright-end excess in the $^{0.1}u$-band fades progressively towards redder 
bands and only disappear completely in the $^{0.1}z$-band.  In this
sense, the $^{0.1}r$-band SDSS LF seems to slightly deviate from this trend. In the 
faint-end, we find a slope that is a bit larger than we
could expect ($\alpha=-1.23$), but this could be just a 
consequence of the fact that the entire SDSS spectroscopic sample
was selected in this band.

\begin{table*}
\begin{center}
\begin{tabular}{cccc}
\hline
\multicolumn{4}{c}{This work} \\
\hline
      Band &      $\Phi_{*}(10^{-2}h^{3}Mpc^{-3})$&        $M_{*}-5log_{10}h$ &       $\alpha$ \\ 
\hline
         $^{0.1}u$ &      $4.16 \pm 0.06$ &      $-17.79 \pm 0.04$  &       $-1.01 \pm 0.03$ \\

         $^{0.1}g$ &      $1.40 \pm 0.06$&        $-19.53 \pm 0.04$&        $-1.10 \pm 0.02$  \\

         $^{0.1}r$ &      $0.90 \pm 0.07$ &        $-20.73 \pm 0.04$ &       $-1.23 \pm 0.02$    \\

         $^{0.1}i$ &      $1.09 \pm 0.05$&        $-20.97 \pm 0.03$&         $-1.16 \pm 0.02$   \\

         $^{0.1}z$ &      $0.89 \pm 0.05$&        $-21.43 \pm 0.04$&         $-1.26 \pm 0.03$ \\

\hline
\hline
\end{tabular}
\begin{tabular}{cccc}
\hline
\multicolumn{4}{c}{Blanton et al. (2003)} \\
\hline
      Band &      $\Phi_{*}(10^{-2}h^{3}Mpc^{-3})$&        $M_{*}-5log_{10}h$ &       $\alpha$ \\
\hline
         $^{0.1}u$ &      $3.05 \pm 0.33$&      $-17.93 \pm 0.03$  &      $ -0.92 \pm 0.07$  \\

         $^{0.1}g$ &      $2.18 \pm 0.08$&        $-19.39 \pm 0.02$&        $-0.89 \pm 0.03$\\

         $^{0.1}r$ &      $1.49 \pm 0.04$&        $-20.44 \pm 0.01$&        $-1.05 \pm 0.01$    \\

         $^{0.1}i$ &      $1.47 \pm 0.04$&        $-20.82 \pm 0.02$ &         $-1.00 \pm 0.02$   \\

         $^{0.1}z$ &      $1.35 \pm 0.04$&        $-21.18 \pm 0.02$&         $-1.08  \pm 0.02$\\

\hline
\hline
\end{tabular}

\end{center}
\caption{Values of Schechter parameters $\alpha$, $M_{*}$ and $\Phi_{*}$ of 
the best-fit Schechter function in all SDSS bands for this work and 
for Blanton et al. (2003b).}
\label{table:LF}
\end{table*}

The bright-end bump  that shows up clearly
in the $^{0.1}u$-band LF (and partially in the $^{0.1}g$-band LF ) is an interesting
discovery that may have implications for our understanding
of galaxy formation and evolution. In order to investigate
the nature of the objects that populate it, we have
selected all galaxies brighter than $-20.5$ in
the $^{0.1}u$-band sample. We will hereafter refer to this sample, which is
comprised of $422$ objects, as the bright-end bump sample (BEBS).

\begin{figure}
\begin{center}
\begin{tabular}{c}
\includegraphics[width=70mm,height=45mm]{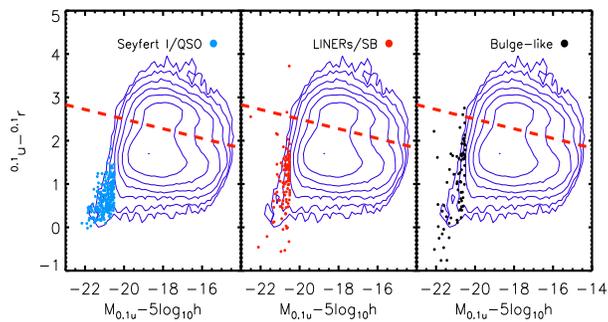}
\end{tabular}
\end{center}
\caption{The $(^{0.1}u-^{0.1}r)$ vs. $M_{^{0.1}u}$ color-magnitude diagram for the three types of bright-end bump galaxies considered: 
Seyferts I/QSO's (left-hand plot), LINER's/SB's (middle plot) and Bulge-like galaxies (right-hand plot). The underlying CMD of the entire $^{0.1}u$-band sample from which BEBS galaxies are selected is shown with log-spaced contours.}
\label{fig:cmd_bb}
\end{figure}

We have visually inspected the spectra of all galaxies in the 
BEBS and classified them into three types. About $60\%$ of objects 
have a typical QSO or Seyfert I spectrum, 
$27\%$ of sources are classified as LINERs or Starburst (SB) and only
$\sim13\%$ of objects are galaxies with a bulge-like spectrum,
and showing little or no signs of star formation. 
In Figure~\ref{fig:cmd_bb} we plot in the $(^{0.1}u-^{0.1}r)$ vs. $M_{^{0.1}u}$ 
colour-magnitude diagram (CMD)
of the $^{0.1}u$-band sample the three types of BEBS galaxies discussed above.
In general, these galaxies form a relatively tight sequence
in the bright-end of the CMD, showing a considerable color
dispersion. Note that, in contrast to Figure~\ref{fig:cmd}, 
density contours are now log-spaced and hence, these objects 
occupy an extremely underpopulated region in the CMD. In the
left-hand plot, QSOs/Seyferts I show the smallest color
dispersion and are, on average, the bluest: $<(^{0.1}u-^{0.1}r)>=0.68$.
Both LINERs/SBs (middle plot) and bulge-like galaxies (right-hand plot)
show much larger color dispersion and are, on average, considerably redder, with 
$<(^{0.1}u-^{0.1}r)>=1.07$ and 
$<(^{0.1}u-^{0.1}r)>=0.91$, respectively. In the $^{0.1}u$-band, bulge-like
galaxies are, on average, the brightest among the BEBS galaxies:   
$<M_{^{0.1}u}>=(-20.97^{QSO/SI},-20.77^{LIN/SB},-21.03^{bulge})$.

We have checked that BEBS galaxies are typically at high
redshift relative to the average redshift of the $^{0.1}u$-band sample 
from which they are drawn. The mean 
redshift in the BEBS is $<z>\sim0.155$ while in the entire $^{0.1}u$-band 
sample is $<z>\sim0.080$.

At this point, it is necessary to remark that this is a 
preliminary analysis. We have performed a rough classification of the 
BEBS galaxies, and this is still subject to some 
uncertainty. However,
our aim is to provide a first approach to the nature of this
population of objects. In Section~\ref{sec:discussion} we speculate on 
possible implications of these results.

\begin{figure}
\plotone{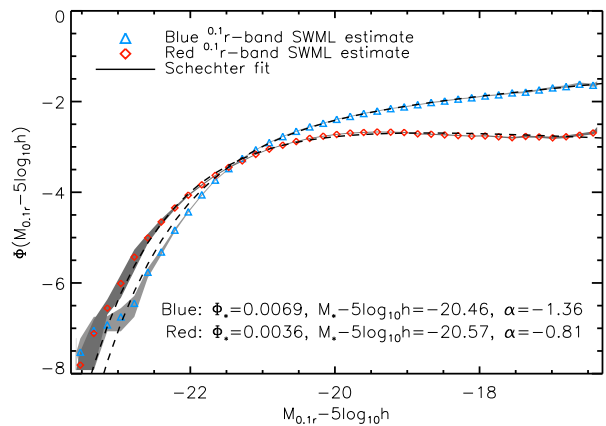}
\caption{The $^{0.1}r$-band SDSS DR6 Luminosity Function for blue and 
red galaxies separately. The SWML LF estimates are shown
in diamonds. The dashed lines represents the best-fit Schechter
function. Best-fit values of Schechter parameters $\alpha$, $M_{*}$ and $\Phi_{*}$ for 
both blue and red galaxies are
also shown in the figure. Shaded regions represent the 1$\sigma$ 
uncertainty calculated using a bootstrapping technique.}
\label{fig:redblue}
\end{figure}

Finally, in Figure~\ref{fig:redblue}, we present, in the same way as in Figure~\ref{fig:lf_r}
and Figure~\ref{fig:ugiz_lf},  SWML estimates of the SDSS $^{0.1}r$-band LF for blue
and red galaxies separately, as well as their corresponding best-fit Schechter LF. Values
of best-fit Schechter parameters are also shown in the figure. The demarcation
that we use to separate blue and red objects is represented by a 
dashed line in Figure~\ref{fig:cmd}. The SDSS $^{0.1}r$-band 
LF of blue galaxies is well fitted by a Schechter LF with $\alpha=-1.36$. In contrast, 
the SDSS $^{0.1}r$-band LF of red galaxies has a negative faint-end slope,
corresponding with $\alpha=-0.81$. These results are consistent with a number of previous
works. In the bright end, at $M_{^{0.1}r}\lesssim-21.5$, the blue 
LF falls remarkably below the red LF.

\subsection{Luminosity Densities}
\label{subsec:LD}

To estimate the luminosity density, $\rho$, in all SDSS bands, we integrate 
throughout the entire absolute magnitude ranges defined by the limits shown
in Table~\ref{table:LD}. In this table we list $\rho$ in all
bands obtained using the SWML estimate of the luminosity functions
shown in Figure~\ref{fig:lf_r} and Figure~\ref{fig:ugiz_lf}. 
Luminosity densities are expressed in $AB~mags~(Mpc^{3}h^{-1})^{-1}$. We 
also give in Table~\ref{table:LD} the effective wavelength corresponding to each SDSS photometric  band, $\lambda_{eff}$.
Errors shown in this table have been calculated using a similar bootstrapping technique
as that discussed in Section~\ref{sec:LF}. We have used $1000$ random sub-samples of $1/3$ the 
number of objects in each sample.

\begin{table*}
\begin{center}
\begin{tabular}{cccccc}
\hline
\hline
      Band &   $\lambda_{eff}$(\AA) &   Absolute Magnitude Range &        $\rho_{SWML}+2.5log_{10}h$ &      $\rho_{Schechter}+2.5log_{10}h$&   $\rho_{Blanton2003}+2.5log_{10}h$\\
\hline
         $^{0.1}u$ &   3216 &  $-23.06 < M_{^{0.1}u} < -14.63$ &       $-14.288 \pm 0.013$ &        $-14.264$&    $-14.10 \pm 0.15$ \\

         $^{0.1}g$ &   4240 &  $-24.93 < M_{^{0.1}g} < -15.63$ &       $-14.910 \pm 0.012$&        $-14.906$&   $-15.18 \pm 0.03$ \\

         $^{0.1}r$ &   5595 &  $-26.03 < M_{^{0.1}r} < -16.16$ &       $-15.769 \pm 0.012$&        $-15.767$&   $-15.90 \pm 0.03$ \\

         $^{0.1}i$ &   6792 &  $-26.23 < M_{^{0.1}i} < -16.63$ &       $-16.113 \pm 0.017$&        $-16.147$ &   $-16.24 \pm 0.03$\\

         $^{0.1}z$ &   8111 &  $-26.53 < M_{^{0.1}z} < -16.93$ &       $-16.500 \pm 0.017$&        $-16.479$&  $-16.56 \pm 0.02$   \\

\hline
\hline
\end{tabular}
\end{center}
\caption{Luminosity densities in $AB~mags~(Mpc^{3}h^{-1})^{-1}$ for all SDSS photometric bands calculated 
using the SWML estimate of the luminosity functions of Figure~\ref{fig:lf_r} and Figure~\ref{fig:ugiz_lf}
($\rho_{SWML}$) and the Schechter LF best fits of Table~\ref{table:LF} ($\rho_{Schechter}$). Absolute magnitude ranges of the 
integration and the effective wavelength corresponding to each band, $\lambda_{eff}$, are also provided. 
 Errors in $\rho_{SWML}$ have been calculated using a bootstrapping technique. In addition, we give for comparison the luminosity densities from Blanton et al. (2003b).}
\label{table:LD}
\end{table*}

As expected, the luminosity density in the nearby universe increases with $\lambda_{eff}$
, in absolute values. This means that the Universe, up to $z\sim0.2$, is considerably
more luminous in the red side of the spectrum than it is in the blue side. Our luminosity densities obtained using the SWML LF estimate are in good agreement with
\cite{Blanton2003a} (see Table~\ref{table:LD}). Both estimates differ in less
than $2\%$ in any band.

However, it is necessary to note that the value of $\rho$ is not too sensitive to
small variations in the shape of the luminosity function. In 
Table~\ref{table:LD}, we also provide luminosity densities in each band obtained using
the Schechter fits to the SWML LF estimates of the luminosity functions discussed in Section~\ref{sec:LF}.
The values of $\rho$ obtained using the SWML estimates and those obtained from
the Schechter fit differ in lees than $1\%$ in each band.

\section{Discussion}
\label{sec:discussion}

The main results presented in this work are the SDSS DR6 Luminosity
Functions of galaxies in the nearby universe. A few years ago,
\cite{Blanton2003a} used an
early version of the
SDSS (DR2) to calculate the SDSS galaxy LFs.
Now, with the SDSS DR6 available, galaxy statistics have
improved by a huge factor of $\sim9$ in the very blue
$^{0.1}u$-band and by a factor of $\sim4-5$ in the rest of the
SDSS photometric bands. Moreover, we have ensured a high redshift 
completeness in our galaxy samples. Firstly, we have 
defined an area on the sky of high angular redshift completeness by 
making use of the survey masks, provided by the NYU-VAGC. 
Secondly, we guarantee that the effect of brightness-dependent 
redshift incompleteness is negligible within the magnitude ranges
that define our galaxy samples. These
advances make our SDSS DR6 LFs substantially more
precise than those from \cite{Blanton2003a} at both
the bright and the faint end. This said, the LFs of
\cite{Blanton2003a}
are compatible with our results. However, notable differences,
which are surely physically significant, exist. In the bright end of the
blue bands LFs (especially in the $^{0.1}u$-band) we find a remarkable
excess, which was very noisy in \cite{Blanton2003a} due to their lack of
statistics. In the
faint end, we obtain steeper slopes in all SDSS bands,
especially in the
$^{0.1}u$-band, where the DR6 statistics allow us to go about $1-1.5$
magnitudes deeper as compared to \cite{Blanton2003a}.
If the huge improvement in the statistics and/or the more
accurate determination of the sample completeness
were not behind this discrepancies, one possible explanation could
come from the so-called evolution correction \citep{Blanton2003a},
which we have not taken into account in the determination of our absolute
magnitudes. Because we
are dealing with relatively nearby
objects, it seems unlikely that this correction could modify our results
significantly.
It is worth mentioning at this point that relatively small variations
in the shape of the LF, which are probably not physically significant -
given the uncertainties we are dealing with - , may translate into
considerable changes in the values of the best-fit Schechter parameters.
It is not convenient, therefore, to make comparisons between different
LFs by just looking at these best-fit Schechter parameters.

We have seen that the bright-end bump that we see in the
blue bands (and marginally in the rest of the bands) is statistically
very significant, according to our standard bootstrapping error
analysis. Moreover, as we mentioned above, we also find clear - but noisy
- evidence of
its existence in \cite{Blanton2003a}. We have also checked that
this excess is not a consequence of any of the limits that we have
imposed to define our samples. This is, therefore, a remarkable result
that may have
strong implications for galaxy evolution.

From a preliminary analysis on the nature of this bright-end bump in the
$^{0.1}u$-band LF,
we have seen
that it is mostly populated by star-forming and active galaxies
($\sim85\%$). The
spectra of these galaxies seem consistent with what we expect
from QSOs/Seyferts I ($\sim60\%$) and LINERs/Starbursts ($\sim25\%$).
It seems, therefore,
that an important fraction of the light that we receive from the
brightest galaxies in the $^{0.1}u$-band would
come from nuclear activity. Only about
$15\%$ of galaxies in bright-end bump seem
to be
bulge-like galaxies. A more detailed study
is needed, however, to fully understand the nature of this bright-end
bump.

The
implications of this new results could be investigated
using semi-analytic models of galaxy formation and evolution (SAMs).
With these models, we can, in principle, evaluate
different processes and feed-back relations that could
reproduce our results.

\section{Summary}
\label{sec:sum}

In this work we make use of the SDSS Sixth Data Release to estimate
the number counts, luminosity functions and luminosity densities of
galaxies
in all SDSS photometric bands. The
SDSS DR6 is, by far, the most complete survey of the nearby universe,
containing redshifts for $\sim 1,000,000$ galaxies down to magnitude
$r\sim17.77$
and covering $\sim7400~ deg^{2}$ on the sky. The huge increase in the
galaxy statistics with respect to the previous work of  \cite{Blanton2003a} and the adequate
treatment of both angular and brightness-dependent redshift completeness in our samples
have allowed us to estimate
the galaxy LFs of the nearby universe with unprecedented accuracy. In
addition, we have calculated,
for the first time, the SDSS galaxy number counts in all photometric
bands using spectroscopic data. Luminosity densities in all SDSS bands
have also been computed.

The main results of this work can be summarized as follows:

\begin{itemize}
 \item The SDSS DR6 galaxy number counts in all SDSS photometric bands are
consistent
 with an Euclidean, not-evolved Universe within a magnitude
 range that is limited by redshift incompleteness
in the
 bright end and by the intrinsic apparent magnitude r-band limit
 of the survey, $r=17.77$, in the faint end.
 \item The SDSS DR6 LF of galaxies in the very blue $^{0.1}u$-band
 deviates considerably from that of \cite{Blanton2003a}. This
 discrepancy can be explained by their lack of statistics,
 which has increased by a factor of $\sim9$ thanks to the SDSS DR6.
 In the faint end, where we can reach about $1$ to $1.5$ magnitudes deeper
 without losing accuracy, our SDSS DR6 LF is considerably steeper than that of
\cite{Blanton2003a}. More interesting,
 in the bright end we find a remarkable excess, of $\sim1.7$ dex at $M_{^{0.1}u}\simeq-20.5$ with
 respect
 to the best-fit Schechter LF. This bright-end bump
 is very strong in the $^{0.1}u$-band and weakens in the $^{0.1}g$-band,
 fading away towards the very red $^{0.1}z$-band.
 \item We conclude that the SDSS DR6 LFs of galaxies in the $^{0.1}g$,
$^{0.1}r$,
$^{0.1}i$
 and $^{0.1}z$ bands are also compatible with
 \cite{Blanton2003a}, considering the large increase in the statistics
thanks to the SDSS DR6,
of a factor
 $\sim4-5$ in these bands. Some significant differences
 exist, however, especially in the faint end, where
 we find a slightly steeper slope and
 we can reach about 1 magnitude deeper without losing precision.
 \item A preliminary analysis of the origin of the bright-end bump
 seen in the $^{0.1}u$-band SDSS DR6 LF reveals that it is
 comprised of QSO and Seyferts I galaxies ($\sim60\%$), Starburst and
LINERs
($\sim25\%$)
 and bulge-like galaxies ($\sim15\%$). It seems,
 therefore, that a big fraction of this exceeding luminosity might
comes from
 nuclear activity.
 \item The $^{0.1}r$-band SDSS DR6 LF of blue galaxies is consistent with a 
 Schechter LF with a remarkably steep faint-end slope, corresponding 
 to $\alpha = -1.36$. The $^{0.1}r$-band SDSS DR6 LF of red galaxies has,
 however, a slightly decreasing faint-end slope, corresponding 
 to $\alpha = -0.81$.   
 \item The SDSS DR6 luminosity densities of galaxies are in very good
 agreement with \cite{Blanton2003a} in all photometric bands, since they
are integrated quantities.
 \end{itemize}

 The state-of-the-art results presented in this paper may be used to
constrain
 a variety of aspects of star formation
 histories or feed-back processes in galaxy formation models. However,
  much effort is still needed in the survey field to fully
 understand the mechanisms that drive the evolution
 of galaxies in the Universe. This is especially necessary at high-z,
 where statistics are still very poor.

\section*{Acknowledgements}

ADMD is supported by the Ministerio de Educaci\'on y Ciencia 
of the Spanish Government  (MEC) through FPI grant AYA2005-07789. ADMD
and FP acknowledge the MEC for their financial support
through PNAYA2005-07789.

We would like to especially thank Michael Blanton for 
providing help on the calculation of the luminosity function
and giving comments on the manuscript. We acknowledge
Michael Blanton and  David Hogg as the corresponding
authors of the NYU-VAGC DR6. 

We also want to thank
Jonatan Hern\'andez-Fern\'andez, V\'ictor M. Mu\~noz Mar\'in, Anatoly Klypin and    
Hans-Walter Rix for stimulating scientific discussions. Finally, we 
thank Antonio J. Cuesta for providing 
technical help throughout the process of making this work.

Funding for the SDSS has been provided by the Alfred P. Sloan
Foundation, the Participating Institutions, NASA, the NSF, 
the U.S. Department of Energy, the Japanese Monbukagakusho, and
the Max Planck Society. The SDSS website is http://www.sdss.org/.

\bibliography{./paper}

\label{lastpage}

\end{document}